# Phase Stability, Structures and Properties of the $(Bi_2)_m \cdot (Bi_2Te_3)_n$ Natural Superlattices

J.-W. G. Bos[1,*], F. Faucheux[1], R. A. Downie[1] and A. Marcinkova[2]

1. Institute of Chemical Sciences and Centre for Advanced Energy Storage and Recovery, Heriot-Watt University, Edinburgh, EH14 4AS, United Kingdom.

2. School of Chemistry and Centre for Science at Extreme Conditions, University of Edinburgh, West Mains Road, Edinburgh, EH9 3JJ, United Kingdom.

[*]j.w.g.bos@hw.ac.uk

The phase stability of the $(Bi_2)_m \cdot (Bi_2Te_3)_n$ natural superlattices has been investigated through the low temperature solid state synthesis of a number of new binary $Bi_xTe_{1-x}$ compositions. Powder X-ray diffraction revealed that an infinitely adaptive series forms for $0.44 \leq x \leq 0.70$, while an unusual 2-phase region with continuously changing compositions is observed for $0.41 \leq x \leq 0.43$. For $x > 0.70$, mixtures of elemental Bi and an almost constant composition $(Bi_2)_m \cdot (Bi_2Te_3)_n$ phase are observed. Rietveld analysis of synchrotron X-ray powder diffraction data collected on $Bi_2Te$ (m = 2, n = 1) revealed substantial interchange of Bi and Te between the $Bi_2$ and $Bi_2Te_3$ blocks, demonstrating that the block compositions are variable. All investigated phase pure compositions are degenerate semiconductors with low residual resistivity ratios and moderate positive magnetoresistances ($R/R_0 = 1.05$ in 9 T). The maximum Seebeck coefficient is $+80\ \mu V\ K^{-1}$ for $x = 0.63$, leading to an estimated thermoelectric figure of merit, $zT = 0.2$ at 250 K.

# 1. Introduction

$Bi_2Te_3$ is among the most widely investigated semiconducting materials due to its excellent thermoelectric properties [1]. Recent developments in nanostructuring have resulted in thermoelectric figures of merit, $zT = 1.5$, exceeding the traditional limit of unity [2, 3]. Elemental bismuth is among the most promising thermoelectric materials for cooling below room temperature [4]. In addition, $Bi_2Te_3$ and Bi are currently attracting much interest as topological insulators [5, 6]. Besides these widely investigated "end-members" an infinitely adaptive series of layered $(Bi_2)_m \cdot (Bi_2Te_3)_n$ natural superlattices consisting of different stacking sequences of Bi double layers and $Bi_2Te_3$ blocks exist [7-10]. These materials are accessed via a low temperature solid state synthesis route and show promising p-type thermoelectric properties near m:n = 2:1 [10]. In addition, pressure induced superconductivity has recently been found in $Bi_4Te_3$ (m = 3, n = 3) [11]. The bismuth tellurides are also of interest as model spintronic materials with ferromagnetism observed in both Mn doped BiTe (m = 1, n = 2) and Mn doped $Bi_2Te_3$ (m = 0, n = 3) [12, 13]. The analogous $(Bi_2)_m \cdot (Bi_2Se_3)_n$ and $(Sb_2)_m \cdot (Sb_2Te_3)_n$ series have also been reported [14-16]. The $Bi_xTe_{1-x}$ compositions reported so far (summarized in [10]) correspond to relatively small numbers of $Bi_2$ and $Bi_2Te_3$ blocks per unit cell (low values of m and n). The aim of this work was to establish the phase stability of these natural superlattice materials through the exploration of more complicated stacking sequences, and compositions close to the end-members. The first reported member of the $(Bi_2)_m \cdot (Bi_2Te_3)_n$ series has m = 1, n = 5 (x = 0.44). The final reported member has m = 15, n = 6 (x = 0.70). For the intermediate compositions ($0.44 \leq x \leq 0.70$) an infinitely adaptive series is postulated to exist [10]. In such a series infinitesimal changes in composition result in fully distinct superstructures, and no two-phase regions are observed. The investigated compositions (given in Table I) were chosen to take into account the unexplored regions of the phase diagram and the potentially useful p-type thermoelectric materials near x = 2/3 [10]. In addition, the crystal structure of $Bi_2Te$ (m = 2, n = 1) was investigated using synchrotron X-ray powder diffraction with the aim of obtaining accurate structural information on this representative member of the $(Bi_2)_m \cdot (Bi_2Te_3)_n$ series.

## 2. Experimental

Polycrystalline $Bi_xTe_{1-x}$ samples were prepared via direct reaction of the elements inside vacuum sealed quartz tubes using a low temperature route. Compositions close to $Bi_2Te_3$ (x = 0.41-0.43) were heated for 3 days at 475 °C, homogenised using a mortar and pestle and heated at 350 °C for 5 days. All other compositions were heated at 250 °C for 10 days with one intermediate homogenisation after 4 days. Trial reactions for x = 0.80 and x = 0.90 at 150 °C resulted in mixtures of Bi and Te. No weight losses were observed in any of the reactions. The phase purity of the prepared materials was investigated by X-ray powder diffraction using a Bruker D8 Advance diffractometer with monochromated Cu $K_{\alpha,1}$ radiation. Lattice constants were obtained from LeBail fits using the JANA2006 program [17]. Room temperature synchrotron X-ray powder diffraction data on $Bi_2Te$ were collected on the ID31 instrument at the European Synchrotron Radiation Facility. The wavelength used was 0.3998 Å. The angular 2θ interval was 2-35° and data were binned in 0.004° steps. Rietveld analysis of this data was performed using the GSAS/EXPGUI suite of programs [18, 19]. A linear absorption correction was used (μR = 2), which did not affect the refined atomic positions. The temperature and field dependence of the electrical resistivity were measured using a Quantum Design Physical Property Measurement System. Contacts were applied in standard four point geometry using silver paint. The dimensions of the bars were ~2 x 2 x 8 mm$^3$. The temperature dependence of the Seebeck coefficient was measured using a homebuilt apparatus.

## 3. Results

*3.1. Phase stability*: The X-ray powder diffraction patterns for the synthesised materials were analysed using the structural model described in our earlier work [10], which is based on that proposed by Lind and Lidin for the analogous Bi-Se series [14]. This model relies on a 4-dimensional superspace description of the crystal structure, which uses a basic hexagonal unit cell with $a \sim 4.4$ Å and $c \sim 6$ Å, and a modulation along the $c$-axis described by a vector $\mathbf{q} = \gamma[001]^*$ in reciprocal space. This description allows for a convenient indexing of the diffraction patterns with a gradual variation of the lattice constants and γ-value as x is varied from 0.4 to 1.0 in $Bi_xTe_{1-x}$. Compositions with rational γ-values can equivalently be described using a conventional unit cell (e.g. $Bi_2Te$) but compositions with irrational γ-values (e.g. BiTe) have incommensurate superstructures, and can only be approximated using 3-dimensional crystallography. The refined lattice constants and γ-values are summarized in Table I. The x-dependence of γ and the cell volume are shown in Fig. 1. For x = 0.41-0.43 the diffraction patterns were fitted using two phases; the first has the $Bi_2Te_3$ structure with γ = 1.20 and freely refined lattice parameters, the second is a $(Bi_2)_m \cdot (Bi_2Te_3)_m$ phase with both the lattice constants and γ freely refined. A comparison of the powder diffraction patterns of these samples clearly reveal the 2-phase nature (Fig. 2). The lattice constants of the $Bi_2Te_3$ and $(Bi_2)_m \cdot (Bi_2Te_3)_m$ phase change gradually (Table I, Fig. 1) revealing that both have varying compositions. This is therefore not a conventional 2-phase region characterised by varying weight fractions of two limiting compositions. Trial LeBail fits keeping the lattice constants for the $Bi_2Te_3$ phase fixed did not adequately fit the observed patterns (e.g. $\chi^2 = 2.2$ for x = 0.42). The phase fractions were estimated using the intensities of the (2-10) reflections (extracted from the LeBail fit) of the phases present (Table I). The thus estimated percentage of the $Bi_2Te_3$ phase gradually decreases from 83% (x = 0.41) to 73% (x = 0.42) to 43% (x = 0.43).

The x = 0.60 and x = 0.63 compositions are single phase (Fig. 2) with lattice constants and γ-values that are consistent with those predicted based on previous work (Fig. 1, Table I). The X-ray diffraction patterns for these samples, however, show substantial broadening compared to $Bi_2Te$ [e.g. the full width at half maximum for the (2-10) reflection is 0.32(1)° compared to 0.24(1)°]. For $Bi_2Te$ it is possible to fit the whole pattern adequately using a single phase. For the two new compositions, the basic cell reflections are well fitted, but the superstructure reflections are not. The predicted width for the (0001) reflection is about 0.25°, while the observed width is 0.50(5)°. This additional broadening may be related to the stacking coherence of the $Bi_2$ and $Bi_2Te_3$ blocks but is not well understood at present. The predicted c-axis lengths are 60 Å and 1200 Å (Table I), which reveals that this broadening is not simply proportional to the number of blocks in the repeat unit. The profile fitting allows the effects of particle size and microstrain to be separated, and revealed that only microstrain contributes to the broadening (of the basic reflections) for these three compositions. More work, including lattice imaging, needs to be done to elucidate the additional superlattice broadening.

For x = 0.73, 0.80 and 0.90, elemental bismuth and a limiting $(Bi_2)_m \cdot (Bi_2Te_3)_n$ phase with almost constant composition are observed. The cell volume and γ-value increase slowly but the magnitude remains close to that observed for the x = 0.7 ($Bi_7Te_3$) phase (Figs. 1, 2, Table I). This reveals that there is a miscibility gap between 0.70 < x < 1.00, where the layered $(Bi_2)_m \cdot (Bi_2Te_3)_m$ phases do not form. The estimated fraction of elemental bismuth increases from 16% for x = 0.73 to 26% for x = 0.80 to 64% for x = 0.90.

*3.2. Crystal structure of $Bi_2Te$*: LeBail fits to the synchrotron X-ray diffraction data confirmed that the structure of $Bi_2Te$ can be described using a simple commensurate superstructure (γ = 4/3). The unit cell contains two $Bi_2$ double layers and a single $Bi_2Te_3$ block, and the structure can be described in the P-3m1 space group. Starting values for the atomic coordinates were obtained from the stacking of ideal $Bi_2$ and $Bi_2Te_3$ blocks. Preliminary Rietveld analysis led to a good agreement between the observed and calculated data ($\chi^2$ = 3.2). However, a large negative temperature factor was found for the Te1 position. This is indicative of Te/Bi site inversion. Refinement of the site occupancies revealed Bi/Te inversion on the Te1 and Bi2 sites with fitted occupancies of 64(2)/36(2) and 68(2)/32(2), respectively ($\chi^2$ = 2.9). These values are within error equal, and were constrained to be the same. The final Rietveld fit is shown in Fig. 3, while the obtained lattice constants, atomic parameters and selected bond lengths are given in Tables II and III. A representation of the crystal structure is shown in Fig. 4. The two Bi double layers are characterised by a short internal bond [Bi1-(Bi/Te)2 = 3.08 Å], and are separated by a long bond [(Bi/Te)2-

[(Bi/Te)2 = 3.56 Å]. These values are similar to those observed in elemental bismuth (3.06 Å and 3.52 Å), which suggests that the (Bi/Te)2 inversion has a limited structural impact. The $Bi_2$ block thickness, however, increases from 1.59 Å (Bi) to 1.68 Å ($Bi_2Te$). This can be explained by a Poisson-type mechanism (volume conservation), in which the compressive strain on the a-axis [10] results in an expansion along the c-direction. The central Bi3-Te2-Bi3 bonds in the $Bi_2Te_3$ block are 3.28 Å, while the outer Bi3-(Te/Bi)1 bonds are 3.14 Å. The equivalent bond distances in $Bi_2Te_3$ are 3.25 Å and 3.07 Å. The modest expansion of the outer Bi-Te bonds (+2 %) could be related to the (Te/Bi)1 site inversion. The net result of volume conservation (a contraction of the $Bi_2Te_3$ block along the c-axis is expected [10]) and the lengthening Bi-Te bond is a modest increase in block thickness from 7.53 Å (for $Bi_2Te_3$) to 7.57 Å (for $Bi_2Te$). The Bi1-(Te/Bi1) bond distance between the $Bi_2$ and $Bi_2Te_3$ blocks is 3.44 Å, which is substantially longer than the weakest bond within the $Bi_2Te_3$ block (3.28 Å) but much shorter than the equivalent distance across the van der Waals gap in $Bi_2Te_3$ (3.65 Å), suggesting that the contraction of the crystallographic c-axis ([10]) is linked to the removal of the van der Waals gaps.

*3.3. Thermoelectric properties*: The temperature dependence of the electrical resistivity is shown in Fig. 5. The x = 0.41-0.43 two-phase samples are degenerate semiconductors with low residual resistivity ratios (RRR ~ 2). This is typical for the layered $(Bi_2)_m \cdot (Bi_2Te_3)_n$ phases but quite different from $Bi_2Te_3$, which has RRR = 22 [10]. The samples become more conducting as the fraction of $Bi_2Te_3$ decreases. The temperature dependence of the Seebeck coefficient of the x = 0.42 sample is shown in Fig. 6. This reveals a crossover from *p*- to *n*-type conduction near 350 K. The small positive value of the Seebeck coefficient in the 100-350 K temperature interval ($S_{290K}$ = +10 µV $K^{-1}$) contrasts with the negative values observed for single phase $Bi_2Te_3$ (x = 0.40; $S_{290K}$ = -175 µV $K^{-1}$) and $Bi_4Te_5$ (x = 0.44; $S_{290K}$ = -30 µV $K^{-1}$) [10]. One possible explanation is that $Bi_2Te_3$ has become p-type due to the replacement of a small fraction of Te by Bi. This scenario is supported by the volume change in Fig. 1, which suggests that some substitution occurs. The resistivity of the x = 0.60 sample shows a broad maximum at 150 K, which is similar to the behaviour observed for $Bi_4Te_3$ (x = 0.57) [10], and signals a crossover from n- to p-type conduction, which is evident in the temperature dependence of the Seebeck coefficient (Fig. 6). The x = 0.63 sample shows typical degenerate semiconducting behaviour and a positive Seebeck coefficient with a broad maximum of 80 µV K$^{-1}$ centred on 250 K. This gives a maximum power factor, PF = 15 µW K$^{-2}$ cm$^{-1}$. The composition dependence of the Seebeck coefficient at 290 K is shown in the inset to Fig. 6, and reveals a plateau at 80-90 µV K$^{-1}$ for $0.63 \leq x \leq 0.70$, and another plateau at -30 µV K$^{-1}$ for $0.44 \leq x \leq 0.57$.

*3.4. Magnetoresistance*: The symmetric part of the magnetoresistance, MR = R/R$_0$, where R$_0$ is the resistance in zero applied field, at 2 K is shown in Fig. 7. The MR for the two-phase samples (x = 0.41-0.43) scales with the $Bi_2Te_3$ content: x = 0.41, MR = 1.23 (83% $Bi_2Te_3$); x = 0.42, MR = 1.19 (73% $Bi_2Te_3$); x = 0.43, MR = 1.09 (42% $Bi_2Te_3$) in an applied field of 9 T. The MR for the single phase x = 0.60 and x = 0.63 compositions is 1.06 and 1.04 in 9 T, respectively. The field dependence of the MR can be fitted to a power law: MR = 1+Ax$^n$ over a wide range of applied fields. In case of the x = 0.41-0.43 samples this expression fits well above 1.5 T (Fig. 7). The exponents are n = 1.25(1) for x = 0.41, n = 1.20(1) for x = 0.42, and n = 1.48(2) for x = 0.43. Below 1.5 T, the MR follows an almost linear field dependence for the mixed phase samples. For x = 0.60 and x = 0.63, the power law captures the entire measured interval satisfactorily (Fig. 7), yielding n = 1.56(1) and n = 1.77(5), respectively. In conventional semiconductors, a small positive MR is expected with a quadratic field dependence in low fields, which changes to a sub-quadratic dependence in larger applied fields. This type of MR is due to the Lorentz force, which leads to a backflow of charge carriers. In most materials this is a small effect (R/R$_0$ < 1.1) but for materials with large carrier mobilities enormous values can be observed (e.g. Bi with R/R$_0$ = 3800 [20]). The magnitude and field dependence of our single phase samples are consistent with normal positive MR. The linear low field dependence for the mixed phase compositions is somewhat unusual. Linear MR has recently been observed in insulating p- and n-type $Bi_2Te_3$ samples (R/R$_0$ ~ 2 in 9T), where it has been linked with topological surface states [21]. Metallic $Bi_2Te_3$ samples have a quadratic low field dependence. However, the linear MR in insulating $Bi_2Te_3$ extends to fields in excess of 9 T, which is far higher than observed in our samples. The cause of the linear low field MR is therefore not clear, and the analysis is complicated by the fact that these are two-phase samples.

## 4. Discussion

The diffraction results show that "predicted" members of the $(Bi_2)_m \cdot (Bi_2Te_3)_n$ series coexist with a $Bi_2Te_3$ phase with varying composition in the 2-phase region for $0.40 < x < 0.44$. The structural parameters for the $(Bi_2)_m \cdot (Bi_2Te_3)_n$ phases are consistent with the established trends for these materials (Fig. 1). The coexistence of two polymorphs with gradually changing structural parameters suggests that they have the same $Bi_xTe_{1-x}$ composition with x equal to the nominal value. This contrasts with the region for $0.70 < x < 1.0$, where two phases with almost fixed composition (Bi and x ~ 0.70) coexist in varying weight fractions. The coexistence of two phases with identical composition is unexpected from thermodynamic considerations and is reminiscent of the $Y_{1-x}Gd_xMnO_3$ series [22]. The end-members can be prepared as hexagonal or orthorhombic polymorphs but not as mixtures. However, for $0.1 < x < 0.5$, hexagonal-orthorhombic mixtures where both phases have identical composition are found, even after prolonged heating at 1200 °C. The only apparent difference between the end-members and the intermediate compositions is the lattice strain resulting from the size-mismatch between $Y^{3+}$ and $Gd^{3+}$. Strain effects due to the size mismatch of the Bi double layers and the $Bi_2Te_3$ blocks play a key role in stabilising the $(Bi_2)_m \cdot (Bi_2Te_3)_n$ infinitely adaptive series [10]. However, it is not clear what effect this has on the unusual phase behaviour observed for $0.41 \leq x \leq 0.43$. The successful synthesis of the m = 5, n = 4 and m = 115, n = 74 phases suggests that a unique superstructure is possible for any composition between $0.44 \leq x \leq 0.70$, and that the series is thus infinitely adaptive in this composition interval. No predicted members of the $(Bi_2)_m \cdot (Bi_2Te_3)_n$ series were found for x = 0.8 and x = 0.9. A phase with composition $Bi_4Te$ (x = 0.8), stable below 120 °C is listed in the published binary phase diagram [9]. Reaction of Bi and Te at 150 °C resulted in mixtures of Bi and Te, which suggests that this phase is not accessible using conventional heating. The full structure refinement for $Bi_2Te$ reveals significant exchange of Bi and Te between the $Bi_2$ and $Bi_2Te_3$ blocks. This demonstrates that the composition of the blocks is an additional variable, in addition to the number of $Bi_2$ and $Bi_2Te_3$ blocks in a repeat unit. The refined bond distances for $Bi_2Te$ reveal that there are only small changes to the bond distances compared to elemental Bi and $Bi_2Te_3$. This is in keeping with the presence of an infinitely adaptive series, where only a small energy stabilisation compared to the end-members, and thus small structural distortions, are expected [15]. The reasons for the atomic site inversion are not clear but it may be that it occurs to minimize the strain from the mismatch between the $Bi_2$ and $Bi_2Te_3$ blocks [10]. The overall stoichiometry is not affected, and neither is the stacking of the blocks ($\gamma$). The facile Bi/Te inversion is unexpected from the tabulated radii (6-fold $Bi^{3+}$ = 1.17 Å , and $Te^{2-}$ = 2.07 Å [23]) but is consistent with high-pressure diffraction studies on $Bi_2Te_3$ and $Bi_4Te_3$ where substitutional alloy formation is observed [11, 24]. In addition, small amounts of $Te_{Bi}$ or $Bi_{Te}$ defects are common at ambient conditions. For example, the $Bi_2Te_3$ sample reported in [10] is n-type due to a slight (sub%) Te excess. A full explanation of the interplay between composition and structure will depend on the accurate determination of a larger number of the $(Bi_2)_m\cdot(Bi_2Te_3)_n$ structures. With regards to the physical properties of these materials; the Bi/Te inversion must lead to a substantial amount of acceptor and donor states near the Fermi level. This may be one reason for the degenerate semiconducting behaviour observed for these samples. The small magnetoresistances, compared to $Bi_2Te_3$ and Bi, could reflect a reduction in charge carrier mobility due to the Bi/Te disorder. A recent report by Sharma et al. gives the thermal conductivity ($\kappa$) of some of these compositions [25]. Addition of $Bi_2$ blocks results in an almost complete suppression of the low temperature phonon peak (T < 100 K) for $Bi_2Te_3$ but has a much smaller effect at higher temperatures. $Bi_2Te$ has $\kappa \sim 2$ W $K^{-1}$ $m^{-1}$ at 250 K. Assuming a similar $\kappa$ for x = 0.63 yields an estimated thermoelectric figure of merit, zT = 0.2 at 250 K.

To conclude, the present results are consistent with the presence of an infinitely adaptive $(Bi_2)_m \cdot (Bi_2Te_3)_n$ series for $Bi_xTe_{1-x}$ compositions between $0.44 \leq x \leq 0.70$. Outside these limits two-phase behaviour is observed. It may be possible to extend this range under kinetic control. The *p*-type conductors near $x = 0.63$ show some promise for thermoelectric refrigeration applications.

**Acknowledgements**

J-WGB acknowledges the EPSRC and the Royal Society for support and the STFC-GB for the beam time at the ESRF.

**Figure captions**

Fig. 1. Composition dependence of (a) the fitted modulation parameter ($\gamma$) and (b) the unit cell volume. The dashed lines indicate the limits of phase stability. The modulation parameter is expected to vary linearly with x. The solid line in the volume plot is a guide to the eye. Black squares are $(Bi_2)_m \cdot (Bi_2Te_3)_n$ members reported before [10], red circles are new additions, while the blue triangles are $Bi_2Te_3$ or Bi. The inset in Fig. 1b shows the x-dependence of the c/a-ratio.

Fig. 2. Composition dependence of the (0001) and (2-10) Bragg reflections for $Bi_xTe_{1-x}$, revealing 2-phase mixtures for $x = 0.41$-$0.43$, pure samples for $x = 0.60$, $x = 0.63$ and $x = 0.67$, and 2-phase mixtures for $x = 0.73$, $x = 0.80$ and $x = 0.90$. The dashed lines are guides to the eye.

Fig. 3. Observed (crosses), calculated (solid line) and difference Rietveld profiles for a fit to synchrotron X-ray powder diffraction data for $Bi_2Te$ ($m = 2$, $n = 1$). The positions of the Bragg reflections are indicated by short vertical lines.

Fig. 4. Schematic representation of the crystal structure of $Bi_2Te$ ($m = 2$, $n = 1$). The labelling of the atoms corresponds to that used in Tables II, III. Dashed and dotted lines indicate weak covalent bonds.

Fig. 5. Temperature dependence of the electrical resistivity for selected $Bi_xTe_{1-x}$ compositions.

Fig. 6. Temperature dependence of the Seebeck coefficient for selected $Bi_xTe_{1-x}$ compositions. The inset show the composition dependence of the Seebeck coefficient at room temperature.

Fig. 7. Field dependence of the magnetoresistance ($R/R_0$) for selected $Bi_xTe_{1-x}$ compositions. The solid lines in (b) are fits to a power law (see text).

Table I. Refined lattice constants and goodness of fit for the investigated $Bi_xTe_{1-x}$ compositions. The phase fraction is estimated from the relative intensities of the (2-10) reflections in the two-phase samples. The number of $Bi_2$ (m) and $Bi_2Te_3$ (n) blocks per 3-dimensional unit cell and the predicted c-axis length are also given.

| Bi fraction | m:n | Formula | c-predicted (Å)[a] | a (Å) | c (Å) | Volume (Å$^3$) | γ | $\chi^2$ | phase fraction |
|---|---|---|---|---|---|---|---|---|---|
| 0.41 | 5:118 | $Bi_{41}Te_{59}$ | 1218 | 4.3812(1) | 6.0931(2) | 101.29(1) | 1.2 | 1.8 | 0.83 |
|  |  |  |  | 4.4007(1) | 6.0370(2) | 101.25(1) | 1.204(1) |  | 0.17 |
| 0.42 | 5:58 | $Bi_{21}Te_{29}$ | 608 | 4.3846(1) | 6.0981(2) | 101.53(1) | 1.2 | 1.5 | 0.73 |
|  |  |  |  | 4.3997(1) | 6.0554(2) | 101.51(1) | 1.209(1) |  | 0.27 |
| 0.43 | 15:114 | $Bi_{43}Te_{57}$ | 405 | 4.3893(1) | 6.0948(2) | 101.69(1) | 1.2 | 1.7 | 0.42 |
|  |  |  |  | 4.4147(1) | 6.0285(2) | 101.75(1) | 1.221(1) |  | 0.58 |
| 0.60 | 5:4 | $Bi_3Te_2$ | 60.0 | 4.4500(3) | 5.9848(4) | 102.64(1) | 1.297(1) | 1.4 | - |
| 0.63 | 115:74 | $Bi_{63}Te_{37}$ | 1196 | 4.4604(2) | 5.9791(4) | 103.02(1) | 1.315(1) | 1.4 | - |
| 0.67 | 2:1 | $Bi_2Te$ | 17.9 | 4.4658(2) | 5.9733(4) | 103.17(1) | 1.333(1) | 1.3 | - |
| 0.73 | 3:1 | $Bi_8Te_3$ | 21.8 | 4.5454(3) | 5.9282(6) | 106.07(2) | 1.5 | 1.5 | 0.16 |
|  |  |  |  | 4.4697(3) | 5.9693(6) | 103.28(2) | 1.338(1) |  | 0.84 |
| 0.80 | 5:1 | $Bi_4Te$ | 29.5 | 4.5455(3) | 5.9314(6) | 106.13(2) | 1.5 | 1.6 | 0.26 |
|  |  |  |  | 4.4734(3) | 5.9645(6) | 103.37(2) | 1.345(1) |  | 0.74 |
| 0.90 | 25:2 | $Bi_9Te$ | 117 | 4.5450(3) | 5.9290(6) | 106.07(2) | 1.5 | 1.7 | 0.62 |
|  |  |  |  | 4.4776(3) | 5.9603(6) | 103.49(2) | 1.357(1) |  | 0.38 |

[a]Calculated from c = 1/3(mc' + nc'') with c' = 11.589 Å and c'' = 30.474 Å.

Table II. Refined atomic parameters for Bi$_2$Te (m = 2, n = 1). The space group is P-3m, lattice constants are a = 4.4688(1) Å, c = 17.9216(4) Å, Goodness of fit parameters: $\chi^2$ = 2.9, wR$_p$ = 9.0%, R$_p$ = 7.1% R$_F^2$ = 5.4%.

|         | Wyckoff | x   | y   | z         | Occ.            | U$_{iso}$ (Å$^2$) |
|---------|---------|-----|-----|-----------|-----------------|-------------------|
| Bi1     | 2c      | 0   | 0   | 0.3383(3) | 0.98(2)         | 0.020(2)          |
| (Bi/Te)2| 2d      | 1/3 | 2/3 | 0.5681(3) | 0.67(2)/0.33(2) | 0.016(2)          |
| Bi3     | 2d      | 1/3 | 2/3 | 0.8880(4) | 1.00            | 0.021(1)          |
| (Te/Bi)1| 2d      | 1/3 | 2/3 | 0.2113(4) | 0.67(2)/0.33(2) | 0.011(2)          |
| Te2     | 1a      | 0   | 0   | 0         | 1.01(2)         | 0.025(4)          |

Table III. Selected bond distances for Bi$_2$Te (m = 2, n = 1).

|                    | distance (Å) |
|--------------------|--------------|
| Bi1-(Bi/Te)2       | 3.077(3)     |
| Bi1-(Te/Bi)1       | 3.435(5)     |
| (Bi/Te)2-(Bi/Te)2  | 3.560(7)     |
| Bi3-(Te/Bi)1       | 3.133(6)     |
| Bi3-Te2            | 3.272(5)     |

Fig. 1a

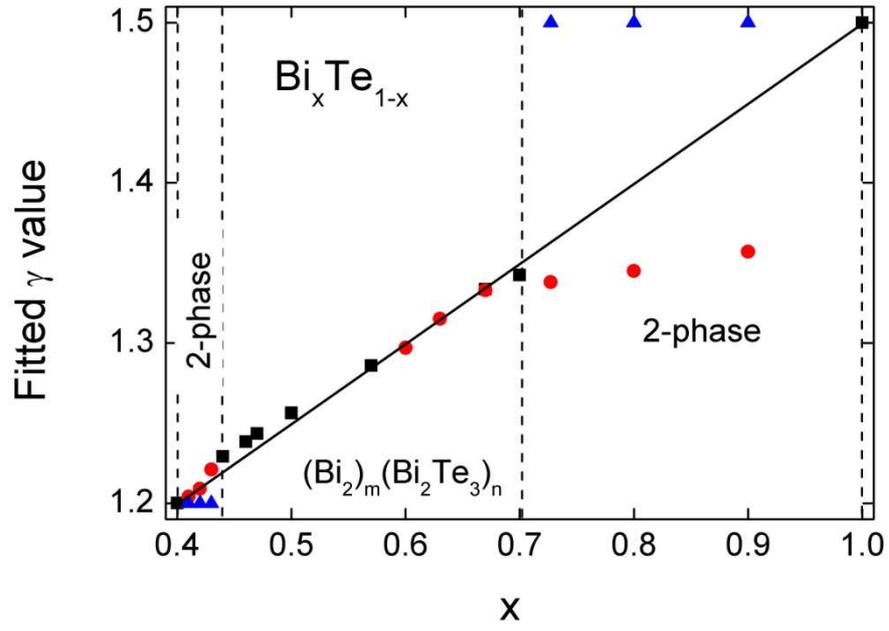

Fig. 1b

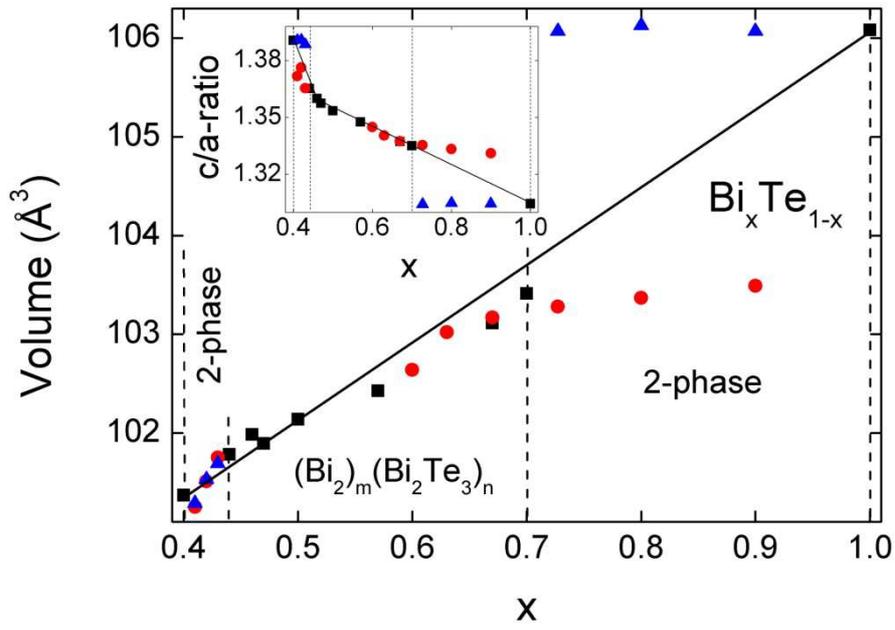

Fig. 2

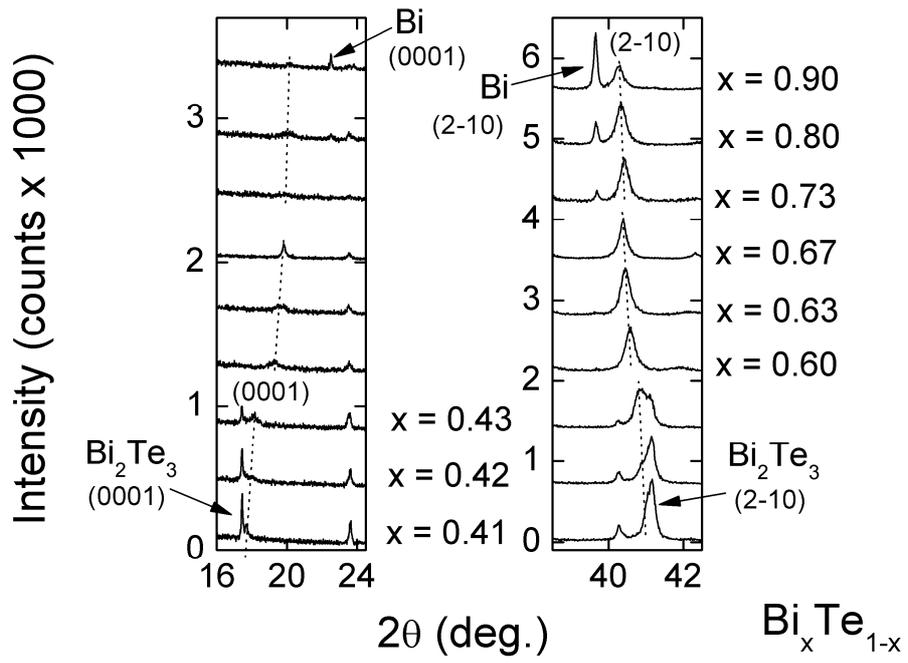

Fig. 3

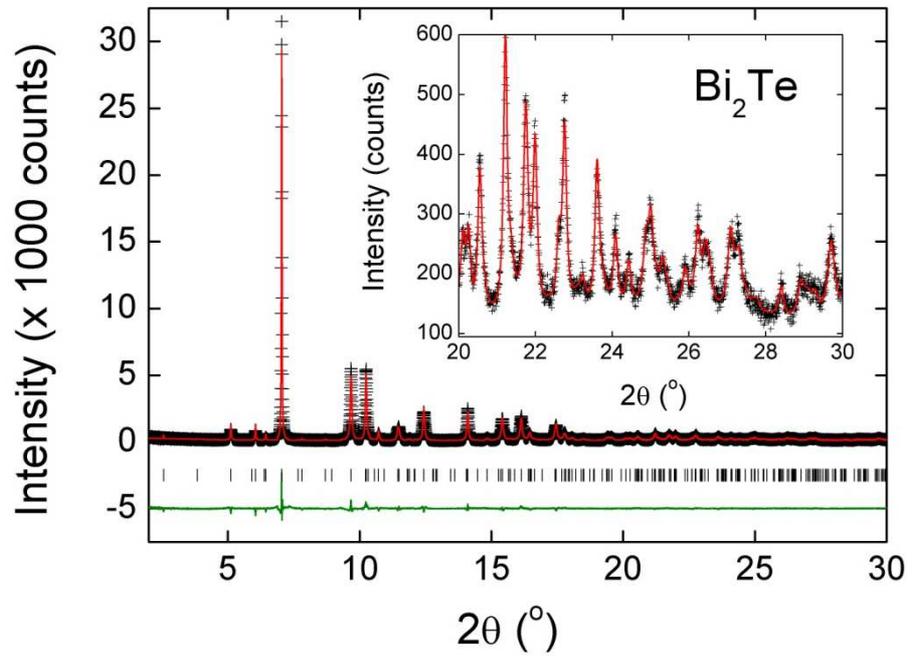

Fig. 4

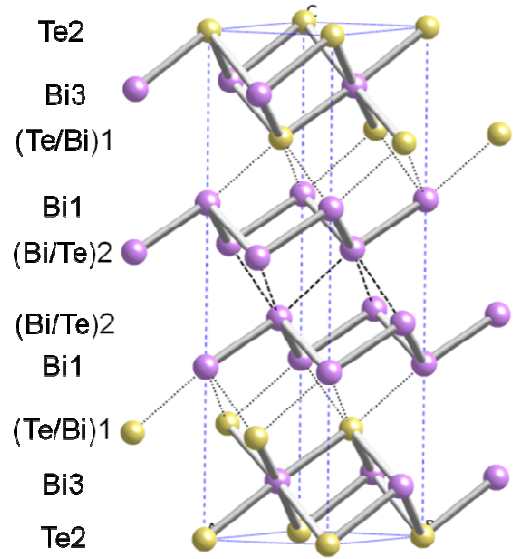

Fig. 5

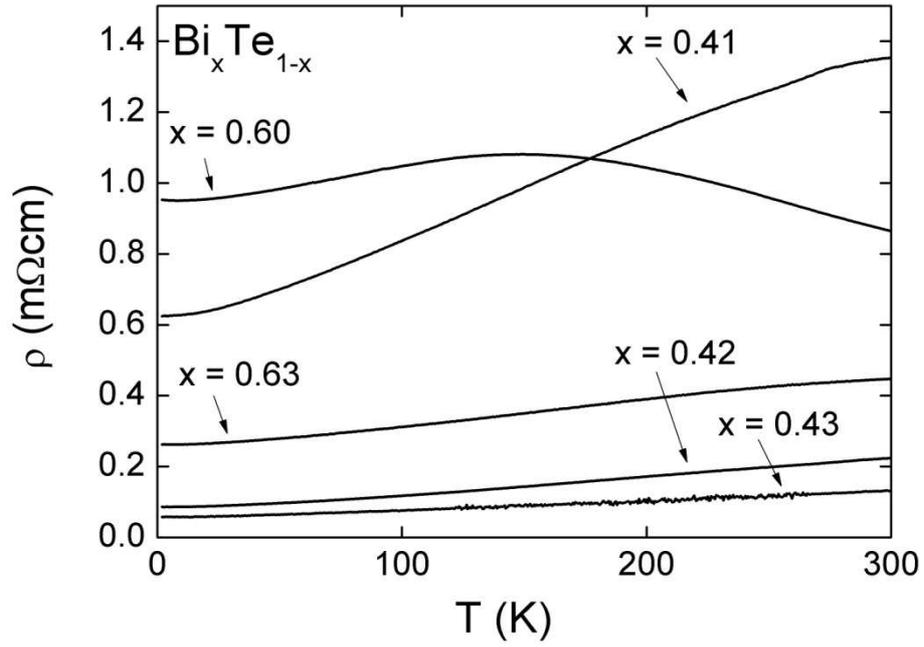

Fig. 6

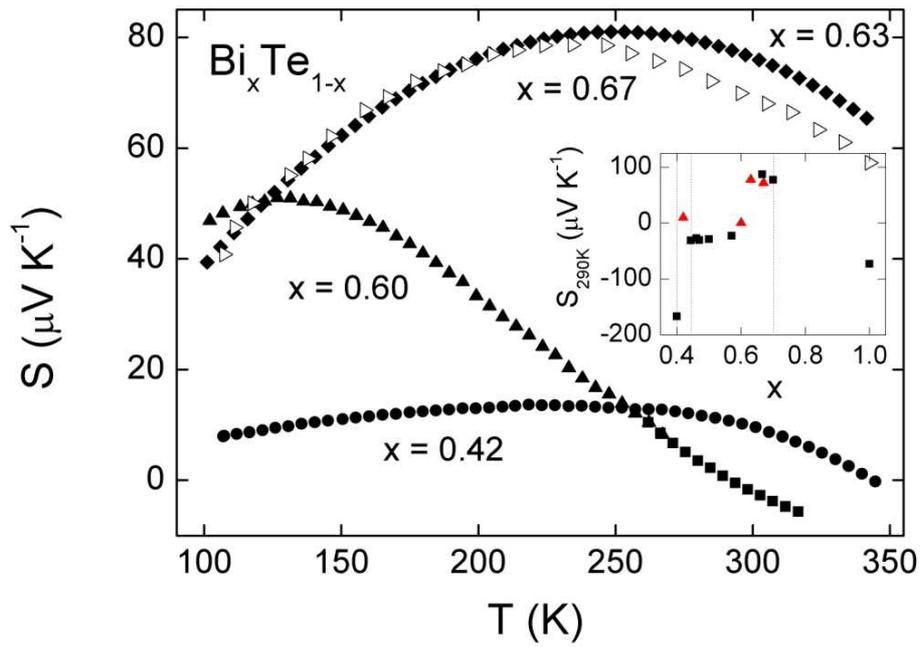

Fig. 7

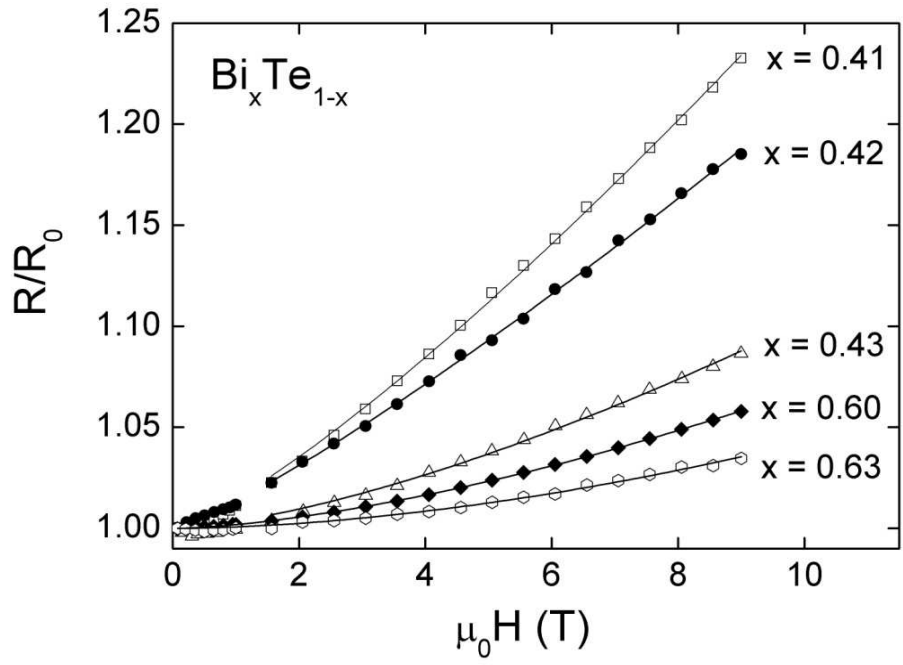